\documentclass[12pt,pre,showpacs,onecolumn,preprint,tightenlines,amsfonts,amsmath,amssymb,byrevtex]{revtex4}
\usepackage{graphicx,pifont}
\usepackage{bm}
\preprint{Submitted to Phys. Rev. E.}
\begin{document}

\markright{\textit{Ball, Dewar, Sugama: 
Metamorphosis of plasma turbulence--shear flow dynamics \ldots}}{} 
\title{Metamorphosis of plasma turbulence--shear flow dynamics through a transcritical 
bifurcation}
\author{R. Ball}
\email{Rowena.Ball@anu.edu.au}
\author{R. L. Dewar}
\affiliation{ Department of Theoretical Physics,
The Australian National University, Canberra ACT 0200 Australia}
\author{H. Sugama}
\affiliation{National Institute For Fusion Science,
Oroshi-cho, Toki GIFU 509-5292 Japan}

\date{June 20, 2002}

\begin{abstract}
The structural properties of an economical model for a confined plasma 
turbulence governor are investigated through bifurcation and  stability 
analyses.
A close relationship is demonstrated between the underlying bifurcation 
framework of the model and typical behavior associated with low- to 
high-confinement transitions such as shear flow stabilization of turbulence 
and oscillatory collective action.
In particular, the analysis evinces two types of discontinuous transition  
that are qualitatively distinct. 
One involves classical hysteresis, governed by viscous dissipation.
The other is intrinsically oscillatory and non-hysteretic, and thus provides 
a model for the so-called dithering transitions that are frequently observed. 
This metamorphosis, or transformation, of the system dynamics is
an important late side-effect of symmetry-breaking, which manifests as 
an unusual non-symmetric transcritical 
bifurcation induced by a significant shear flow drive.  
\end{abstract}

\pacs{52.30.-q, 52.25.Xz, 05.45.-a, 52.35.Ra, 0240.Xx}

\maketitle

\section{\label{one}}

Fusion plasmas, and possibly other quasi two-dimensional fluid systems, 
may undergo a more-or-less dramatic transition from a low to a 
high confinement state (the L--H transition) as the power input is increased, 
with the desirable outcome that particle and energy confinement is greatly 
improved due to localized transport reduction
\cite{Terry:2000}. 
In this work we report on a bifurcation and stability probe of an 
economical model for L--H transition dynamics that uncovers a 
mechanism by which a radical change, or metamorphosis, may occur in the 
qualitative nature of the dynamics. 
We apply the results of this analysis to clarify the relationship between 
the structure of the model and the physics of the process that it describes, 
and draw comparisons with characteristics of L--H transitions observed in various experiments. 

Since 1988 there has been much progress in developing
low-dimensional (low-order or reduced) descriptions of L--H transition dynamics and
associated oscillatory phenomena (see, for example, Refs 
\onlinecite{Itoh:1988,Hinton:1991,Diamond:1994,%
Pogutse:1994,Voj:1995,Sugama:1995,Lebedev:1995,%
Drake:1996,Hu:1997,Takayama:1998,Itoh:1998,%
Staebler:1999,Thyagaraja:1999,Ball:2001}), the driving force being the potential power of 
a unified, low-dimensional model as a predictive tool for the 
design and control of confinement states.
For example, a model that speaks of
the shape and extent of hysteresis in the L--H transition would help 
engineers who are interested in controlling access to H-mode. Given the 
many
variables and parameters that {\em could} be varied around a hysteretic 
r\'egime, it would be cheaper---i.e., save hundreds of cpu hours and/ or
many expensive diagnostics---to know in advance which ones actually {\em do} 
affect the hysteresis, and which do not. 

To help construe the context in which low-dimensional descriptions
of plasma dynamics
are sought, it is appropriate at this stage to make some general remarks.
It makes sense to try to find the simplest description of an evolving system
that is consistent with the time and space scales on which one is interested in making
experimental observations of that system. One would like the description to
incorporate the qualitative nature of the system structure and dynamics, so that
it can be used for design and control purposes
 and make useful predictions. A truly useful description
usually turns out to be a low-dimensional system of coupled ordinary
differential equations.
Such descriptions are powerful because they are supported
by well-developed theories that give qualitative and global insight, such as
bifurcation, stability, and symmetry theory
\cite{Golubitsky:1985,Holmes:1996}.
In principle we can map analytically the bifurcation structure of the entire 
state and parameter space of a low-dimensional dynamical system, but this is
not yet possible for an infinite-dimensional system.

However, the quest for a low-dimensional state space that captures 
the qualitative dynamics of L--H transitions has been problematic. 
It has been shown \cite{Ball:2000,Ball:2001} that some of the 
models cited above do not reflect salient features of L--H
transitions such as shear flow suppression of turbulence, or
are incomplete, or show profound structural discrepancies, 
although it is intuitively reasonable to expect that 
manifestly different models should be equivalent at some deeper level if they 
describe the same phenomena.

By economical, or minimal, model we mean the smallest, functionally
simplest, and mathematically consistent model that captures qualitatively the
dynamical traits that are typically observed over many experiments in
different machines. The strength and potency of a minimal model is just this
universality; its apparent disregard for details, numbers and unit dimensions  
is sometimes perceived---wrongly---as a weakness.
In keeping with this ideology we introduce here a consensus dynamical model
that is economical in terms of variables and parameters, and
incorporates the smallest number of rate processes of simplest
functional form needed to reflect the universally observed dynamics.
If the model is successful we expect additional terms to have
only quantitative, not qualitative or structural, effects.
We should also be able to identify easily its limits of
validity, or where it breaks down and why.

In section \ref{two} we introduce the plasma turbulence governor 
as a useful schema to conceptualize and  represent the major contributing 
rate and feedback processes, relating these
to the corresponding dynamical system. Bifurcation and stability analyses 
and interpretive discussions, with reference to reported experiments,
 are given in the remaining sections.
In section \ref{three} we begin by determining
the two highest order (most degenerate) singularities in the system,
or {\em organizing centers}. Section \ref{four} describes the generic
bifurcation diagram and discusses the hysteresis and limit cycles
in the system.  In section \ref{five} we illustrate and discuss the useful
properties of the two-parameter bifurcation diagram. 
This discussion leads in to section \ref{six}, in which
we determine explicitly the transcritical metamorphosis 
to an oscillatory, non-hysteretic r\'egime. A short summary is given
in section \ref{seven}.
The Appendix contains a derivation of the dynamical equations.

\section{\label{two} } 

\begin{figure}
\includegraphics[scale=1]{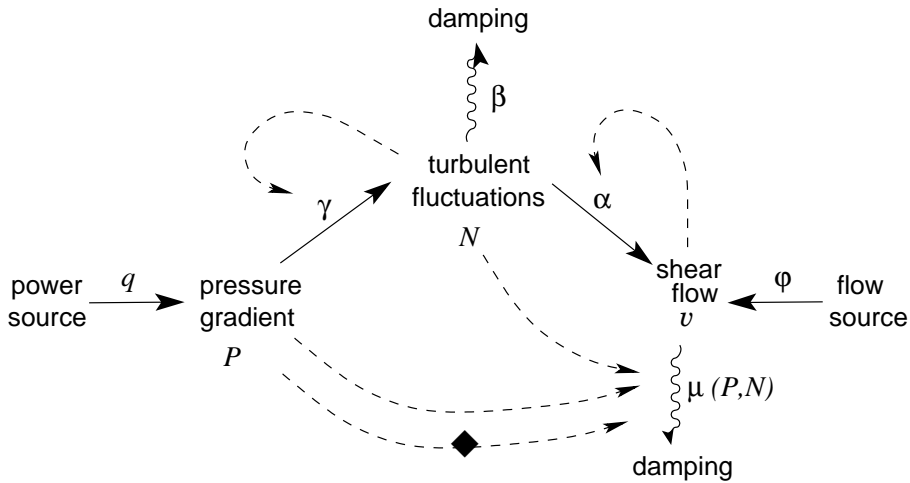}
\caption{\label{fig1}Coupled rates and 
feedback processes that contribute to the dynamics of L--H transitions. 
Solid arrows indicate generation rates, wavy
arrows dissipation, 
dashed arrows feedback 
on rate coefficients; black diamond indicates negative feedback.
}
\end{figure}

The schematic in Fig. \ref{fig1} is a primitive of a plasma turbulence 
governor. (The name is intended to refer to
archetypical mechanical exemplars of feedback controllers such as James Watt's
1788 steam-engine governor. In \cite{James:1987} a comparable scheme was called 
the ``barotropic governor'',
in the context of quasi two-dimensional atmospheric flows.)
A~power input $q$ creates a pressure 
gradient $P$ from which the turbulent density fluctuation intensity $N$
grows at a rate with coefficient~$\gamma$. 
The turbulence feeds energy into the poloidal 
shear flow $v$ via the 
Reynolds stress~$\alpha$. The shear flow is generated externally 
at rate $\varphi$ and damped by the ion viscosity $\mu$. The turbulence
is damped quadratically with coefficient $\beta$.  
Also indicated is a {\em competitive} distribution of energy from
the pressure gradient, whereby different fractions may partition
into turbulence generation and shear flow damping. 
It is not difficult to appreciate how the various rate and competitive 
processes in Fig. \ref{fig1} could balance out---or rather,
{\em un-}balance out---so as to give rise to the oscillatory and hysteretic
dynamics that are characteristic of L--H transitions.

The reduced dynamical system that models this scheme is based on the 
Sugama-Horton model \cite{Sugama:1995}, which itself was derived from 
approximate resistive MHD vorticity and pressure convection equations
\cite{Strauss:1977,Strauss:1980}: 
\begin{align}
\varepsilon\frac{dP}{dt} = &q - \gamma P N\label{dp}\\[2mm] 
\frac{dN}{dt} = & \gamma P N - \alpha v^2N - \beta N^2\label{dn}\\[2mm]
2\frac{dv}{dt} = & \alpha v N - \mu(P,N)v + \varphi\label{dv}\\[2mm]
&\mu(P,N) = bP^m + a P^rN \label{mu}.
\end{align}
In terms of the shear flow kinetic energy $F=v^2$ Eqs \ref{dn} and \ref{dv} may be
written as 
\begin{align}
\frac{dN}{dt} = & \gamma P N - \alpha FN - \beta N^2
\tag{\ref{dn}$^\prime$}\\[2mm]
\frac{dF}{dt} = & \alpha F N - \mu(P,N)F + \varphi F^{1/2}\tag{\ref{dv}$^\prime$}.
\end{align}
The derivation of this system is given in the Appendix. 
The most important modification to the original Sugama-Horton model 
is the symmetry-breaking term $\varphi$ in Eq. \ref{dv}.  
It will be seen that this term, which may be interpreted as an external shear flow 
driving rate, has dramatic effects on the bifurcation structure of the system. 

The first and second terms in the bipartite viscosity function, Eq. \ref{mu}, 
model the neoclassical and anomalous viscosity damping respectively. 
In a plasma of low collisionality the exponent $m$ is 
negative so a high pressure gradient has the effect of
blocking the neoclassical contribution. (Refer to Fig. \ref{fig1}.)
Under these circumstances 
energy can accumulate in the shear flow then 
feed back into turbulence decorrelation. 
On the other hand, a high pressure gradient and
high turbulence levels both {\em enhance} the anomalous viscosity damping,
because the exponent $r$ is positive. The net effect will depend on the
relativity of the three competitive rates involved in the distribution
of energy from the pressure gradient. 

\section{\label{three}}

Generally in bifurcation analysis we are interested in the multiplicity, 
stability, singularity, and parameter dependence of zero solutions of a 
bifurcation equation 
$g=G(x,\lambda_1,\lambda_2,\ldots,\lambda_n)$, 
where $x$ is
a state variable and the $\lambda_i$ are parameters, that is derivable
(in principle if not always in practice) from the equilibria of a 
dynamical system.

In Eqs \ref{dp}--\ref{mu} we may select $x\equiv P$ and the
principal bifurcation parameter $\lambda_1\equiv q$  and set the 
right hand sides to zero to obtain the bifurcation equation,
\begin{equation}\label{g}
g=\frac{1}{2P^2\alpha\gamma^2}\left(aP^rq - q\alpha 
+ bP^{1+m}\gamma\right)
\left(q\beta - P^2\gamma^2\right) + 
\frac{\varphi\left(P^2\gamma^2-q\beta\right)^{1/2}}
{2\left(P\alpha\gamma\right)^{1/2}}
\end{equation}
(where Eq. \ref{dv}$^\prime$ has been used). 
Singular points occur where $g=g_P=0$. (Subscripts on $g$ denote partial 
derivatives with respect to the subscripted variable.) 
On the $v=0$ branch they are given by
\begin{equation}\notag
(P,q,\varphi)=\left(P_i,P_i^2\gamma^2/\beta,0\right),
\end{equation}
with the $P_i$ given by the real, positive roots of $ \beta b+P^{1-m}(aP^r
-\alpha)\gamma = 0$. At these points $g_q=0$ and 
$g_{PP}= -8\left(aP_i^r\left(-1-m+r\right)
+\left(1+m\right)\alpha\right)\gamma^2/\left(\alpha\beta\right). $
Thus for some values of the exponents $m$ and $r$ 
one or more of the singularities may comply with the pitchfork conditions
\begin{equation}\label{pf}
g=g_P=g_q=0=g_{PP}= 0, \,g_{PPP}\neq 0,\, g_{Pq}\neq 0.
\end{equation}
Obviously (since $g_{PP}$ must equal 0), 
compliance with these conditions also implies the existence of hysteresis.

To specify the dependence of the viscosity damping on the pressure gradient in 
Eq. \ref{mu} we set $m=-3/2$ and $r=1$, as in \cite{Sugama:1995}. 
This value of $m$ applies for the temperature dependence of the ion viscosity in a
low collisional r\'egime \cite{Leontovich:1965}. The value of $r=1$ is
the simplest that is consistent with the suggested dependence of the 
anomalous viscosity on the ion temperature in \cite{Sugama:1994a}. 

\begin{figure}
\hspace*{-0.5cm}
	\includegraphics[scale=0.65]{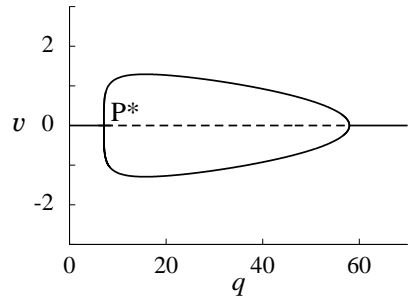} 
	\caption{Bifurcation diagram for the critical set (P*), 
$\varphi=0$, $b=18.58$, $\alpha=2.4$, $\beta=1$, $\gamma=1$, 
$a=0.3$, $\varepsilon = 1.5 $.}\label{figure2}
\end{figure}

With this specification the conditions in Eq. \ref{pf} applied to Eq. \ref{g} 
find the unique pitchfork~P* as 
\begin{equation}\tag{P*}
\left(v,q,b,\varphi\right)=
\left(0,\frac{\alpha^2\gamma^2}{9a^2\beta},
\frac{2\alpha^3\gamma\sqrt{\alpha/a}}{27\sqrt{3}\:a^2\beta},0
\right).
\end{equation}
At P* the two non-degeneracy conditions in Eq. \ref{pf} 
evaluate as $g_{Pq}=8a/\alpha$,
$g_{PPP}= -18a\gamma^2/(\alpha\beta)$.
A pitchfork is described as a 
codimension 2 singularity, because its universal unfolding requires 
2 parameters additional to the principal bifurcation parameter. 
Note that the second unfolding parameter, chosen here as $b$, can be
any of the dissipative parameters $a$, $b$, or $\beta$.
For reference the bifurcation diagram in Fig. \ref{figure2} has been computed
and plotted for the critical set (P*). The singular point 
on the $v=0$ branch at high $q$ complies with the conditions
\begin{equation}\label{tr} 
g=g_P=g_q=0,\, g_{PP}\neq 0, \,\det d^2g<0, 
\end{equation}
where $d^2g$ is the Hessian matrix of second partial derivatives
$
\left(
\begin{array}{cc}
\!g_{PP}& \!g_{q P}\\
g_{q P}& g_{qq}
\end{array}\!\right).
$
A singular point that satisfies these conditions is usually termed 
a transcritical bifurcation, or sometimes a ``simple bifurcation''. 

For non-critical values of $b$ (i.e., $b\neq b_{\left({\text P}^*\right)}=18.58\ldots$), 
P* collapses to a 
second transcritical bifurcation on the $v=0$ branch. These two 
transcriticals coalesce and annihilate each other 
at a second codimension 2 singularity D* on the $v=0$ branch, defined 
by the conditions
\begin{equation}\label{D}
g=g_P=g_q=\det d^2g=0,\, g_{PP}\neq 0,\, g_{Pq}\neq 0,
\end{equation}
and found using Eq. \ref{g} as
\begin{equation}\tag{D*}
 (v,q,b,\varphi)= \left(0,\frac{\left(5\alpha\gamma\right)^2}
{\left(7a\right)^2\beta},
\frac{50\sqrt{5\alpha/\left(7a\right)}\,\alpha^3\gamma}{7^3a^2\beta},0\right),
\end{equation}
with $g_{PP}=-32\gamma^2/\left(7\beta\right)$, $g_{Pq}=16a/(5\alpha)$. 
The bifurcation diagram showing this point 
(at $(v,q,b,\varphi)=(0,0.61\ldots,53.52\ldots,0)$  for 
values of the
other parameters as in Fig. \ref{figure2}) would be extremely dull and 
flat---it consists only of the line $v=0$. 
Such a highly dissipative system has no interesting behaviour at all.

\section{\label{four} }

\begin{figure}
\hspace*{-1cm}
	\includegraphics[scale=0.9]{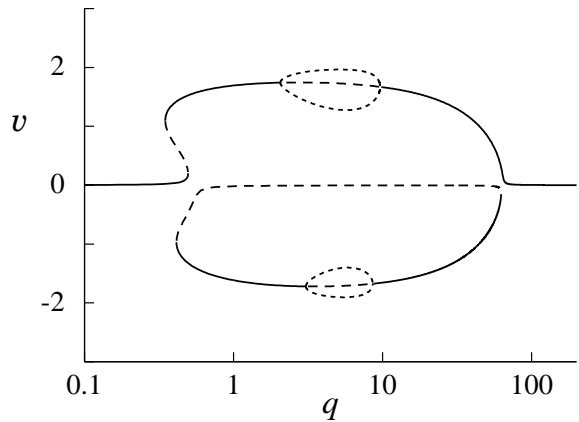} 
\vspace*{-1mm}		
\caption{Bifurcation diagram with  
 $b=1$, $\varphi=0.02$, other parameters as for Fig. \ref{figure2}. 
}\label{figure3}
\end{figure}

A bifurcation diagram for non-critical values of the unfolding parameters 
$b$ and $\varphi$ is shown in Fig. \ref{figure3}.
(In the bifurcation diagrams 
stable solution branches are indicated by solid lines, unstable 
branches by dashed lines, and the dotted lines trace out the maximum and minimum 
amplitude of  limit cycle branches.)
The symmetry evident in Fig.~\ref{figure2} is broken 
by selection of a small positive value of $\varphi$, which determines
a preferred direction of the poloidal shear flow. 

Branches of stable limit cycles
emanate from Hopf bifurcations on the $+v$ and $-v$ H-mode branches.
They reflect reports from experiments that a transition to a quiescent H-mode can 
be achieved followed by the onset of oscillatory behaviour,
or edge-localised modes (ELMs), 
as the power input continues to be increased 
\cite{Ida:1990,Thomas:1998a,Igitkhanov:1998,Shats:1999}. 
 The original Sugama-Horton model was found to exhibit 
a chaotic time series for a particular set of parameter
values in this r\'egime \cite{PEBak:2001}. In our model we have found that this branch of
limit cycles can undergo several successive period doublings followed by period halvings
back to a period-one limit cycle. 

\begin{figure}\hspace*{-0.5cm}
\vbox{
\includegraphics[scale=0.85]{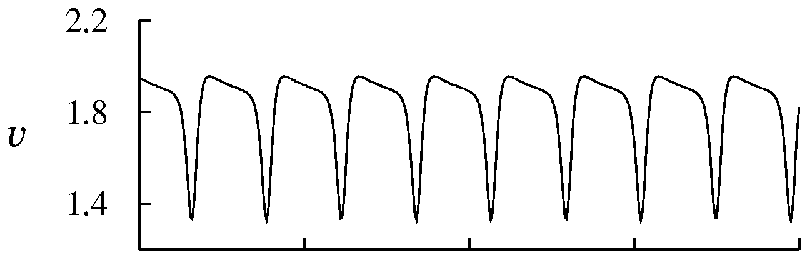}\vspace*{-8mm}\\
\includegraphics[scale=0.85]{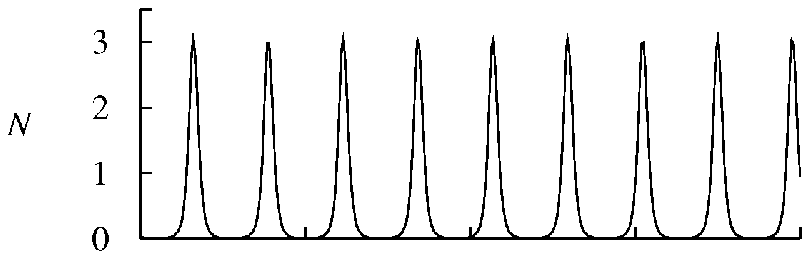}\vspace*{-8mm}\\
\hspace{3mm}\includegraphics[scale=0.84]{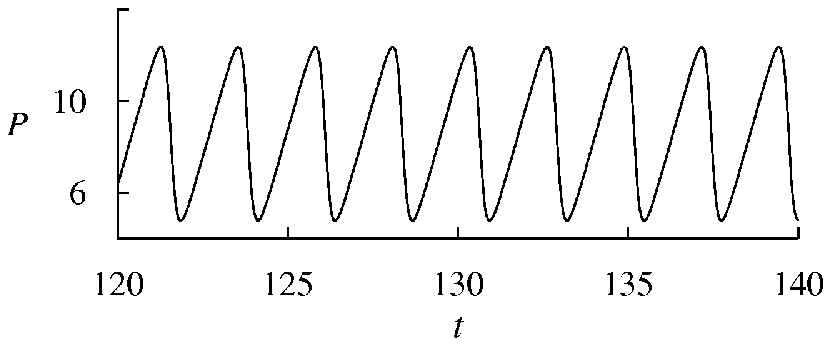}
}	
\caption{A time series for $q=4$ on the $+v$ branch.
 Other parameters are the same as in 
Fig. \ref{figure3}. }\label{figure4}
\end{figure}

The limit cycles are also {\em extinguished} at Hopf bifurcations. 
In \cite{Ball:2001} it was shown that oscillatory behaviour is
regulated by the contribution of the pressure gradient evolution.
At moderately high power input the pressure gradient and turbulence are high, 
neoclassical viscous damping is inhibited, and 
large amplitude oscillations would be expected as energy
alternately accumulates
in the shear flow and is exchanged with the turbulence. 
The relative phases of $v$, $N$, and $P$
are shown in the time series of Fig.~\ref{figure4}.
However, this is
balanced by the {\em enhancement} of {\em anomalous} viscosity damping
by the larger amounts of turbulence and pressure gradient
energy at higher power inputs.
(Refer to the governor schematized in Fig. \ref{fig1}.) 
As this anomalous viscosity effect begins
to take over the amplitude of the limit cycles decreases rapidly until
they are extinguished at the Hopf bifurcations at higher $q$. 

Although definitive experiments have not yet been performed 
that measure the growth and extent of the
H-mode oscillations over the power input, 
it is physically reasonable that they would
be limited by some damping factor. 
The passage through an oscillatory r\'{e}gime with increasing power 
is a characteristic of type III ELMs \cite{Connor:1998,Suttrop:2000}.
However, the quantitative features of type III ELMs, such as
the frequency spectrum, are not reproduced by this simple model.

On the $v<0$ branch the limit cycles are smaller in amplitude and
occur over a smaller range of the power input. At $q\approx 2.5$, 
for example, the $+v$ H-mode is oscillatory but the $-v$ H-mode is
quiescent. Again, to our knowledge the appropriate experiments 
have not yet been carried out, but this is reminiscent of the 
prescription given in ref.
\cite{Burrell:2001}: ``The key factors in creating the quiescent H-mode 
operation
are neutral beam injection in the direction opposite to the plasma current 
(counterinjection) plus cryopumping to reduce the density.'' 

Reports of reversals in the direction of main or impurity ion poloidal 
shear flow \cite{Bell:1998,Solomon:2001} can
also be rationalized on the basis of Fig. \ref{figure3}. In a system 
that is evolved initially 
onto the $v<0$ branch, the poloidal shear 
flow {\em must} reverse if a perturbation
decreases the power input slightly below that at the lower limit point. 
Shear flow reversal
may also occur anywhere along the $v<0$ branch, if the system is given a 
sufficiently strong transient kick.

Note that in Fig. \ref{figure3} the shear flow $v$ reaches a broad 
maximum with increasing power input, then
decreases to the the pre-transition level given by the shear flow source. 
This would be 
reasonable behaviour on physical grounds---one would not expect the shear 
flow to increase indefinitely with power input, because the turbulent 
viscosity damping (the second term in Eq. \ref{mu} with $r=1$)
begins to take over as the power input increases the pressure gradient. 

Clearly there is scope for tuning other parameters in the model so as to 
obtain a complete picture of the steady states and limit cycles over parameter 
space, and more quantitative agreements with experiments.
One may wish, for example, to maximize the range of $q$ over which 
turbulence stabilization
occurs, or minimize the range of $q$ over which limit cycles occur,
or both. 

Figure \ref{figure3} also shows the hysteresis that is predicted by 
compliance with the conditions in Eq. \ref{pf}. 
Transitions with hysteresis have been observed in several machines:
DIII-D \cite{Thomas:1998a},
Asdex Upgrade \cite{Zohm:1995,Ryter:1998}, JET, and in simulations of
ITER \cite{Igitkhanov:1998},
and Alcator C-Mod \cite{Hubbard:1998}. Hysteresis is typically 
modified by dissipation,  
characterised in this model by the parameters $\beta$, $b$, and $a$. 
However, hysteresis does not seem to be a necessary or universal feature of 
discontinuous transitions. 
 
\begin{figure}
\hspace*{-1cm}
\includegraphics[scale=0.65]{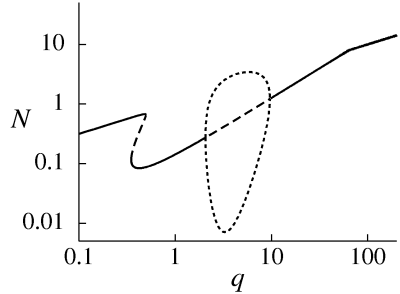}
\vspace*{-1mm}		
\caption{Bifurcation diagram with $N$ as chosen state variable, 
showing the curve that corresponds to the positive $v$ branch in 
Fig. \ref{figure3}. }\label{figure5}
\end{figure}

One of the typical features of L--H transitions that a minimal model
should reflect is suppression of the turbulence by the shear flow. 
Figure \ref{figure5} shows the bifurcation diagram with 
the mean square turbulence level $N$
as the state variable, where for clarity only the curve that matches the
positive $v$ branch is given. The turbulence 
is clearly suppressed over the hysteretic region,
then begins to grow again as the higher pressure gradient 
from higher power input creates more turbulence. 

\section{\label{five}}

\begin{figure} 
\includegraphics[scale=0.85]{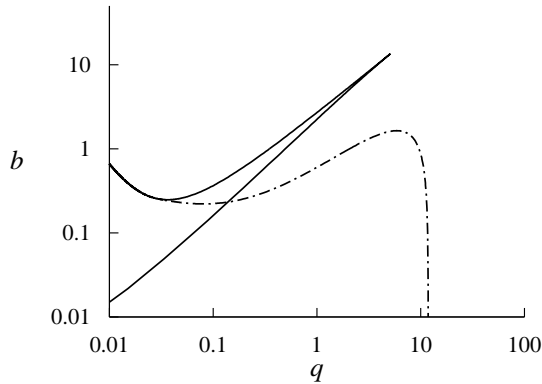}
\caption{Two-parameter curves of the singular points 
in Fig.~\ref{figure3}. Solid lines are the loci of the limit points, dot-dash
lines are the Hopf bifurcation loci. 
}\label{figure6}
\end{figure}

The width and extent of hysteresis for selected values of $b$
can be judged from the {\em two-parameter bifurcation diagram} 
for the $+v$ branch in Fig. \ref{figure6}, in which   
computed curves of the singular points in Fig. \ref{figure3} are shown. 
The solid lines mark the loci of limit points (which are also sometimes
called fold or saddle-node bifurcations) as $b$ is varied. 
The dot-dash line is the locus of Hopf bifurcations as $b$ is varied. 
If one can imagine taking slices across this 
diagram at various important values of $b$, the
bifurcation story of the system can be told compactly, by inferring 
a reconstruction of the
single-parameter $(q,v)$ bifurcation diagram corresponding to each selected 
value of $b$. 

A slice taken above the critical value of $b$ at the cusp would 
yield a $(q,v)$ bifurcation diagram that shows no 
multiplicity of states. Thus, in a highly dissipative system any 
transition is expected to be smooth and gradual rather than discontinuous, 
and a number of experiments suggest this conjecture.
In ASDEX Upgrade the power hysteresis disappears at higher density (which
implies more collisional damping) where gradual 
rather than discontinuous confinement improvement 
occurs \cite{Ryter:1998}. A r\'egime in which density fluctuation 
amplitudes are reduced continuously was also observed in 
\cite{Moyer:1999}.
In \cite{Dahi:1998} a discontinuous bifurcation of the electric field 
in a stellarator was reported for conditions of low neutral density, 
where the charge-exchange damping rate is low. The change in the
electric field became gradual for conditions of high neutral 
density, because the charge-exchange damping rate increases. 
(The electric field is related to the poloidal shear flow and the
pressure gradient through the radial force balance 
\cite{Shats:1997a}.)

Oscillatory behaviour is also expected to be damped out at high dissipation 
rates. The maximum in the locus of Hopf bifurcations in Fig. \ref{figure6} 
occurs at the value of $b$ where the two Hopf bifurcations on
the $+v$ branch in Fig. \ref{figure3} annihilate each other
(or conversely, are created). Above this value of $b$ the
$+v$ branch is stable with no associated limit cycles.

As slices are taken at lower $b$ the hysteresis and the range of oscillatory
behavior evidently become broader.
At low dissipation rates the feedback is strong and
nonlinear behavior is expected to be more pronounced. 

The crossing of the Hopf and limit point loci in Fig.~\ref{figure6}
is non-local, i.e., 
the value of $P$ (and of $v$ and $N$) at the crossing 
on the Hopf curve is different
from that on the limit point curve. Within the overlapping region a 
direct transition to an oscillatory H-mode may occur. 

\section{\label{six}}
\begin{figure}
\hspace*{-5mm}\hbox{
\includegraphics[scale=0.7]{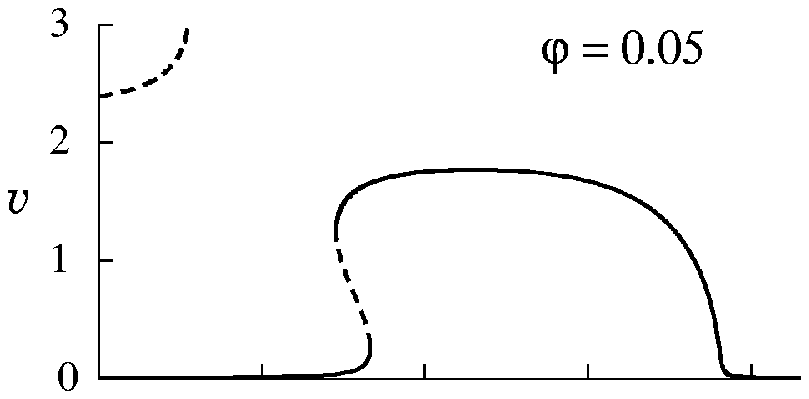}
\hspace{-1cm}\includegraphics[scale=0.7]{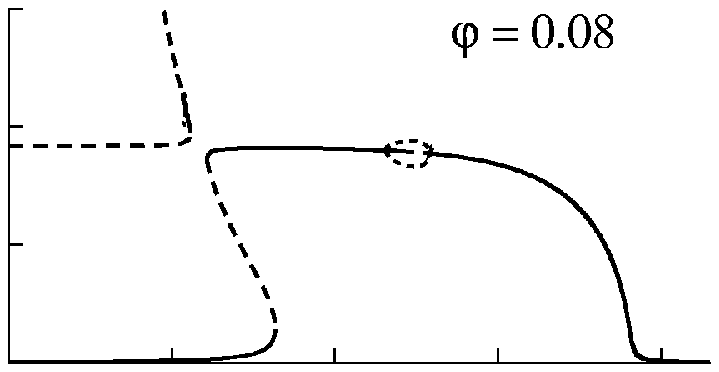}
\hspace{-1cm}\includegraphics[scale=0.7]{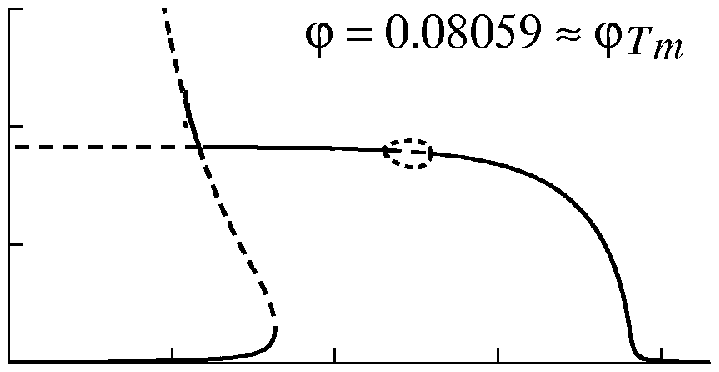}}	
\vspace{-0.5cm}
\hspace*{-5mm}\hbox{
\includegraphics[scale=0.7]{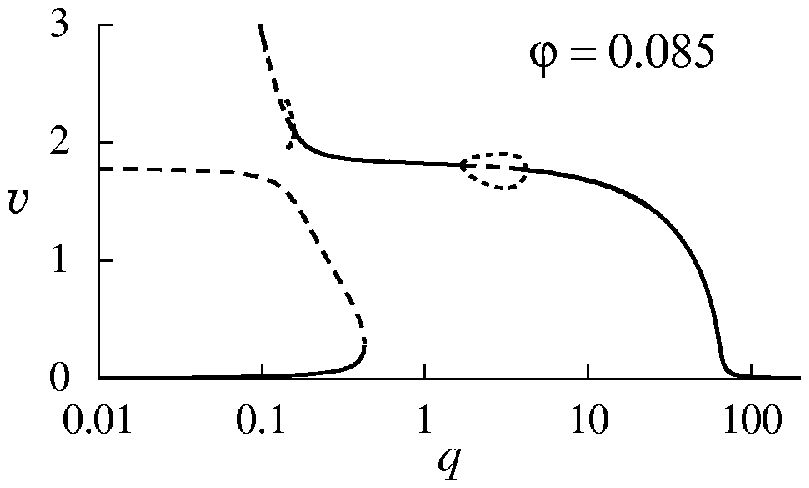}
\hspace{-1cm}\includegraphics[scale=0.7]{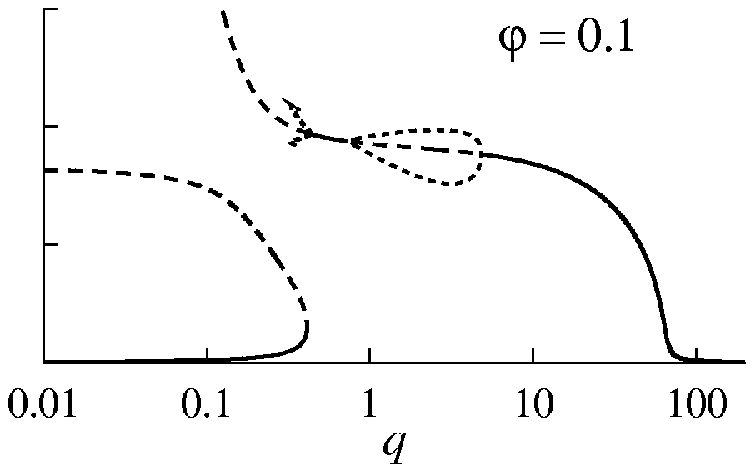}
\hspace{-1cm}\includegraphics[scale=0.7]{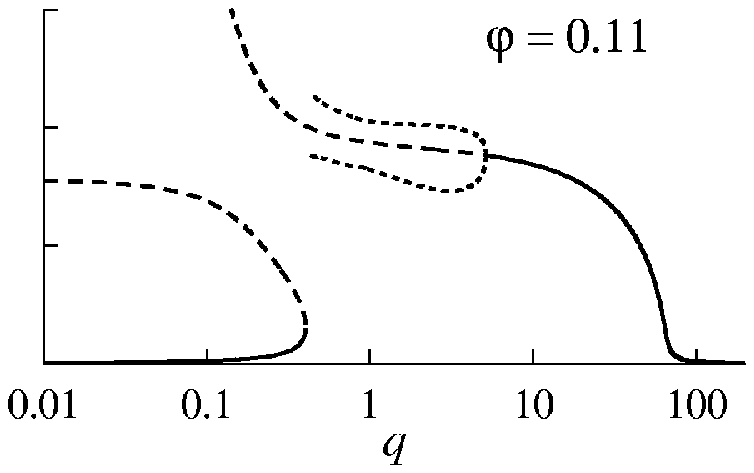}}
\caption{
A series of bifurcation diagram snapshots taken at increasing values of
$\varphi$ illustrates the exchange at $\varphi_{T_m}$
and its aftermath. 
Here $\varepsilon=1.0$ and other parameters are the 
same as in Fig. \ref{figure3}. }	
\label{figure7}
\end{figure}

The minimum in the limit point curve of Fig. \ref{figure6} implies
the existence of another transcritical bifurcation (defined by Eq. \ref{tr})
in the system, that occurs at non-zero~$v$. 
This non-symmetric transcritical bifurcation 
may have important issues concerning access to and 
control of confinement states. Consider the
series of stills in Fig. \ref{figure7}, which  
snapshot  $+v$ bifurcation diagram sections for 
increasing values of $\varphi$. 

As $\varphi$ is incremented a separate branch of solutions to Eq. \ref{g}, 
which is trapped at $(q,v)=(0,\infty)$
for $\varphi=0$, begins to intrude more prominently into the physical 
region ($\varphi=0.05 $). Although it is unstable at first, and therefore physically irrelevant
for very small values of $\varphi$, it does not remain so. The singular occurrence of
a zero real eigenvalue and a pair of complex conjugate eigenvalues with zero
real part signals the appearance of a degenerate Hopf bifurcation. 
(Eigenvalues were computed numerically.) 

Further increments in $\varphi$ separate the limit point and the 
Hopf bifurcation, between which the solutions are stable ($\varphi=0.08 $). 
The branch of limit cycles that emanates from 
the Hopf bifurcation undergoes one or more period doubling bifurcations 
before ending, presumably at a homoclinic (infinite period) terminus. 
This branch of limit cycles is quite short, and thus not very well resolved
in Fig. \ref{figure7} for the lower values of $\varphi$. Note also that the 
branch of limit cycles emanating from the hysteretic solution branch has
also appeared by $\varphi=0.08 $. 

At a metamorphic value of $\varphi$ that we designate $\varphi_T{_m}$ 
the arms of the two separate steady-state branches are exchanged at a 
transcritical bifurcation. 
We know this point is present because the defining conditions Eq. \ref{tr} 
are satisfied with $\varphi\neq 0$. 
(In numbers
$(v,q,\varphi)_T{_m}=(1.8247..,0.1468..,0.08059..)$, with 
$\det d^2g=-0.004687..$, $g_{PP}=-0.001250..$, for values of the other 
parameters
as given in Fig \ref{figure3}. Note that the value of $\varepsilon$ is 
irrelevant for calculating steady-state bifurcations such as the 
pitchfork and transcritical but {\em not} for Hopf bifurcations.)

After the exchange, e.g. at $\varphi=0.085$ and $\varphi=0.1$, 
we see the unusual
occurrence of {\em three} Hopf bifurcations on the same branch, although this
situation could be inferred from the shallow but distinct minimum and the 
maximum in the
locus of Hopf bifurcations in Fig. \ref{figure6}. 

The last frame of Fig. \ref{figure7}, for $\varphi=0.11$, is taken after 
 the ``new'' and the lower-$q$ ``old'' Hopf bifurcations have collided and 
annihilated 
each other at a singular point associated with two zero eigenvalues. 
This is what the minimum in the curve of Hopf bifurcations means. 
The branch of limit cycles emanating from the upper-$q$ ``old'' Hopf bifurcation 
now continues to the (presumed) homoclinic terminus. There are a couple of
period-doublings on it (not shown). 
We also see that the limit cycles are extinguished and
 a quiescent H-mode is achieved at the single remaining Hopf bifurcation. 

Turning our attention to the stable part of the lower branch 
in the last frame of Fig.~\ref{figure7}, we see that
as $q$ is tuned past the lower limit point 
the system  must jump to another stable attractor. This 
transition is {\em very} different from the intrinsically hysteretic transition 
depicted in Fig \ref{figure3}. 
Here the stable attractor on the upper branch is a limit cycle 
rather than a fixed point. 
Furthermore, this transition is {\em not} hysteretic. In fact, hysteresis
is (locally) forbidden by the condition $g_{PP}\neq 0$ of Eq. \ref{tr}. 
Therefore it is not modulated by dissipation in the same way as 
the transition in Fig.\ref{figure3}, although the 
feedback itself is still due to nonlinear dissipation rates. 

As the value of $\varphi$ is increased even further, bifurcation diagrams 
that one could plot gradually become
less meaningful. This is because $\varphi=$ constant is a 
first approximation, valid for small $\varphi$, to to a nonlinear function 
$\varphi(\zeta)$, where 
 $\zeta$ may include dynamical variables and parameters. 

This type of transition could serve as a model for  the
dithering or L--H--L transitions, followed by a quiescent H-mode,
that have been reported in many machines. 
Although there may be other mechanisms for dithering transitions---another
 possible scenario is given at the end of section \ref{five} and indicated in Fig. 
\ref{figure6}, where an oscillatory transition may occur in a very 
poorly dissipative system---we have at least a preliminary 
semiological and classification guide:
if your transition is oscillatory and
non-hysteretic then perhaps you should look for a strong shear flow source,
 if it is strongly hysteretic perhaps you should look at dissipation 
mechanisms.   
Some experimental evidence supports the idea that dithering 
transitions result from a strong shear flow source. 
In \cite{Hugill:1998}, an analysis of time series data around the L--H transition in
 COMPASS-D suggested that a homoclinic orbit is involved in the change of stability
at the transition.  In stellarator W7-AS typically the quiescent ELM-free H-mode is 
obtained after a phase characterized by quasi-periodic ELMs 
\cite{Hirsch:1998,Hirsch:2000}.
In H1 stellarator a transition to fluctuating H-mode occurs at lower gas filling 
pressures and lower magnetic fields than the transition to quiescent H-mode 
\cite{Shats:1997b}. 

In terms of the governor in Fig. \ref{fig1} a shear flow that is generated internally
and driven externally at comparable rates is likely to give rise to 
interesting non-linear dynamics, because more kinetic energy in the shear
flow leads to more turbulence suppression through decorrelation, but also a larger
damping effect, which then alters the competitive distribution of energy from
the pressure gradient. 

\section{\label{seven}}\enlargethispage{1cm}
In summary, this reduced dynamical model, comprised simply of
energy input, exchange, and loss rates,  reflects 
generic characteristics of confined plasma bulk dynamics 
that have not been reflected in previous models. 
The bifurcation and stability analysis also 
reveals two qualitatively different 
transitions. 
The hysteretic transition is controlled by the damping rate coefficients. 
The non-hysteretic transition occurs when there is a relatively strong 
shear flow drive. 

Symmetry-breaking in this system has two major effects.
Firstly, a non-zero shear flow drive is physically inevitable,
even in the best-controlled experiments, and it determines a preferred
direction for the shear flow. Secondly, it interacts with the internal
generation and loss dynamics to 
cause the metamorphosis shown in Fig. \ref{figure7}.

More generally, the information obtained from this analysis strengthens
 the thesis developed in \cite{Holmes:1996}: that
remarkably low-dimensional
models can capture and help explain essential aspects of
turbulent flows that elude understanding from numerical simulations
that include resolved spatial scales, and that physical deductions can
be made from observations of bifurcations.

\begin{acknowledgments}
We thank J. Frederiksen for bringing to our attention the barotropic
governor in ref. \cite{James:1987}. R.B. would like to thank the
Australian Research Council for financial support. 
\end{acknowledgments}

\bibliographystyle{/users/theophysA/rxb105/latex/revtex4/apsrev}

\begin{thebibliography}{47}
\expandafter\ifx\csname natexlab\endcsname\relax\def\natexlab#1{#1}\fi
\expandafter\ifx\csname bibnamefont\endcsname\relax
  \def\bibnamefont#1{#1}\fi
\expandafter\ifx\csname bibfnamefont\endcsname\relax
  \def\bibfnamefont#1{#1}\fi
\expandafter\ifx\csname citenamefont\endcsname\relax
  \def\citenamefont#1{#1}\fi
\expandafter\ifx\csname url\endcsname\relax
  \def\url#1{\texttt{#1}}\fi
\expandafter\ifx\csname urlprefix\endcsname\relax\def\urlprefix{URL }\fi
\providecommand{\bibinfo}[2]{#2}
\providecommand{\eprint}[2][]{\url{#2}}

\bibitem[{\citenamefont{Terry}(2000)}]{Terry:2000}
\bibinfo{author}{\bibfnamefont{P.~W.} \bibnamefont{Terry}},
  \bibinfo{journal}{Reviews of Modern Physics} \textbf{\bibinfo{volume}{72}},
  \bibinfo{pages}{109} (\bibinfo{year}{2000}).

\bibitem[{\citenamefont{Itoh and Itoh}(1988)}]{Itoh:1988}
\bibinfo{author}{\bibfnamefont{S.-I.} \bibnamefont{Itoh}} \bibnamefont{and}
  \bibinfo{author}{\bibfnamefont{K.}~\bibnamefont{Itoh}},
  \bibinfo{journal}{Phys. Rev. Lett.} \textbf{\bibinfo{volume}{60}},
  \bibinfo{pages}{2276} (\bibinfo{year}{1988}).

\bibitem[{\citenamefont{Hinton}(1991)}]{Hinton:1991}
\bibinfo{author}{\bibfnamefont{F.~L.} \bibnamefont{Hinton}},
  \bibinfo{journal}{Phys. Fluids B} \textbf{\bibinfo{volume}{3}},
  \bibinfo{pages}{696} (\bibinfo{year}{1991}).

\bibitem[{\citenamefont{Diamond et~al.}(1994)\citenamefont{Diamond, Liang,
  Carreras, and Terry}}]{Diamond:1994}
\bibinfo{author}{\bibfnamefont{P.~H.} \bibnamefont{Diamond}},
  \bibinfo{author}{\bibfnamefont{Y.~M.} \bibnamefont{Liang}},
  \bibinfo{author}{\bibfnamefont{B.~A.} \bibnamefont{Carreras}},
  \bibnamefont{and} \bibinfo{author}{\bibfnamefont{P.~W.} \bibnamefont{Terry}},
  \bibinfo{journal}{Phys. Rev. Lett.} \textbf{\bibinfo{volume}{72}},
  \bibinfo{pages}{2565} (\bibinfo{year}{1994}).

\bibitem[{\citenamefont{Pogutse et~al.}(1994)\citenamefont{Pogutse, Kerner,
  Gribkov, Bazdenkov, and Ossipenko}}]{Pogutse:1994}
\bibinfo{author}{\bibfnamefont{O.}~\bibnamefont{Pogutse}},
  \bibinfo{author}{\bibfnamefont{W.}~\bibnamefont{Kerner}},
  \bibinfo{author}{\bibfnamefont{V.}~\bibnamefont{Gribkov}},
  \bibinfo{author}{\bibfnamefont{S.}~\bibnamefont{Bazdenkov}},
  \bibnamefont{and}
  \bibinfo{author}{\bibfnamefont{M.}~\bibnamefont{Ossipenko}},
  \bibinfo{journal}{Plasma Phys. Control. Fusion}
  \textbf{\bibinfo{volume}{36}}, \bibinfo{pages}{1963} (\bibinfo{year}{1994}).

\bibitem[{\citenamefont{Vojtsekhovich et~al.}(1995)\citenamefont{Vojtsekhovich,
  Dnestrovskij, and Parail}}]{Voj:1995}
\bibinfo{author}{\bibfnamefont{I.~A.} \bibnamefont{Vojtsekhovich}},
  \bibinfo{author}{\bibfnamefont{A.~Y.} \bibnamefont{Dnestrovskij}},
  \bibnamefont{and} \bibinfo{author}{\bibfnamefont{V.~V.}
  \bibnamefont{Parail}}, \bibinfo{journal}{Nuclear Fusion}
  \textbf{\bibinfo{volume}{35}}, \bibinfo{pages}{631} (\bibinfo{year}{1995}).

\bibitem[{\citenamefont{Sugama and Horton}(1995)}]{Sugama:1995}
\bibinfo{author}{\bibfnamefont{H.}~\bibnamefont{Sugama}} \bibnamefont{and}
  \bibinfo{author}{\bibfnamefont{W.}~\bibnamefont{Horton}},
  \bibinfo{journal}{Plasma Phys. Control. Fusion}
  \textbf{\bibinfo{volume}{37}}, \bibinfo{pages}{345} (\bibinfo{year}{1995}).

\bibitem[{\citenamefont{Lebedev et~al.}(1995)\citenamefont{Lebedev, Diamond,
  Gruzinova, and Carreras}}]{Lebedev:1995}
\bibinfo{author}{\bibfnamefont{V.~B.} \bibnamefont{Lebedev}},
  \bibinfo{author}{\bibfnamefont{P.~H.} \bibnamefont{Diamond}},
  \bibinfo{author}{\bibfnamefont{I.}~\bibnamefont{Gruzinova}},
  \bibnamefont{and} \bibinfo{author}{\bibfnamefont{B.~A.}
  \bibnamefont{Carreras}}, \bibinfo{journal}{Phys. Plasmas}
  \textbf{\bibinfo{volume}{2}}, \bibinfo{pages}{3345} (\bibinfo{year}{1995}).

\bibitem[{\citenamefont{Drake et~al.}(1996)\citenamefont{Drake, Lau, Guzdar,
  Hassam, Novakovski, Rogers, and Zeiler}}]{Drake:1996}
\bibinfo{author}{\bibfnamefont{J.~F.} \bibnamefont{Drake}},
  \bibinfo{author}{\bibfnamefont{Y.~T.} \bibnamefont{Lau}},
  \bibinfo{author}{\bibfnamefont{P.~N.} \bibnamefont{Guzdar}},
  \bibinfo{author}{\bibfnamefont{A.~B.} \bibnamefont{Hassam}},
  \bibinfo{author}{\bibfnamefont{S.~V.} \bibnamefont{Novakovski}},
  \bibinfo{author}{\bibfnamefont{B.}~\bibnamefont{Rogers}}, \bibnamefont{and}
  \bibinfo{author}{\bibfnamefont{A.}~\bibnamefont{Zeiler}},
  \bibinfo{journal}{Phys. Rev. Lett.} \textbf{\bibinfo{volume}{77}},
  \bibinfo{pages}{494} (\bibinfo{year}{1996}).

\bibitem[{\citenamefont{Hu and Horton}(1997)}]{Hu:1997}
\bibinfo{author}{\bibfnamefont{G.}~\bibnamefont{Hu}} \bibnamefont{and}
  \bibinfo{author}{\bibfnamefont{W.}~\bibnamefont{Horton}},
  \bibinfo{journal}{Phys. Plasmas} \textbf{\bibinfo{volume}{4}},
  \bibinfo{pages}{3262} (\bibinfo{year}{1997}).

\bibitem[{\citenamefont{Takayama et~al.}(1998)\citenamefont{Takayama, Unemura,
  and Wakatani}}]{Takayama:1998}
\bibinfo{author}{\bibfnamefont{A.}~\bibnamefont{Takayama}},
  \bibinfo{author}{\bibfnamefont{T.}~\bibnamefont{Unemura}}, \bibnamefont{and}
  \bibinfo{author}{\bibfnamefont{M.}~\bibnamefont{Wakatani}},
  \bibinfo{journal}{Plasma Phys. Control. Fusion}
  \textbf{\bibinfo{volume}{40}}, \bibinfo{pages}{775} (\bibinfo{year}{1998}).

\bibitem[{\citenamefont{Itoh et~al.}(1998)\citenamefont{Itoh, Toda, Yagi, Itoh,
  and Fukuyama}}]{Itoh:1998}
\bibinfo{author}{\bibfnamefont{S.-I.} \bibnamefont{Itoh}},
  \bibinfo{author}{\bibfnamefont{S.}~\bibnamefont{Toda}},
  \bibinfo{author}{\bibfnamefont{M.}~\bibnamefont{Yagi}},
  \bibinfo{author}{\bibfnamefont{K.}~\bibnamefont{Itoh}}, \bibnamefont{and}
  \bibinfo{author}{\bibfnamefont{A.}~\bibnamefont{Fukuyama}},
  \bibinfo{journal}{Plasma Phys. Control. Fusion}
  \textbf{\bibinfo{volume}{40}}, \bibinfo{pages}{737} (\bibinfo{year}{1998}).

\bibitem[{\citenamefont{Staebler}(1999)}]{Staebler:1999}
\bibinfo{author}{\bibfnamefont{G.~M.} \bibnamefont{Staebler}},
  \bibinfo{journal}{Nuclear Fusion} \textbf{\bibinfo{volume}{39}},
  \bibinfo{pages}{815} (\bibinfo{year}{1999}).

\bibitem[{\citenamefont{Thyagaraja et~al.}(1999)\citenamefont{Thyagaraja, Haas,
  and Harvey}}]{Thyagaraja:1999}
\bibinfo{author}{\bibfnamefont{A.}~\bibnamefont{Thyagaraja}},
  \bibinfo{author}{\bibfnamefont{F.~A.} \bibnamefont{Haas}}, \bibnamefont{and}
  \bibinfo{author}{\bibfnamefont{D.~J.} \bibnamefont{Harvey}},
  \bibinfo{journal}{Phys. Plasmas} \textbf{\bibinfo{volume}{6}},
  \bibinfo{pages}{2380} (\bibinfo{year}{1999}).

\bibitem[{\citenamefont{Ball and Dewar}(2001)}]{Ball:2001}
\bibinfo{author}{\bibfnamefont{R.}~\bibnamefont{Ball}} \bibnamefont{and}
  \bibinfo{author}{\bibfnamefont{R.~L.} \bibnamefont{Dewar}},
  \bibinfo{journal}{J. Plasma Fus. Res.} \textbf{\bibinfo{volume}{4}},
  \bibinfo{pages}{266} (\bibinfo{year}{2001}).

\bibitem[{\citenamefont{Golubitsky and Schaeffer}(1985)}]{Golubitsky:1985}
\bibinfo{author}{\bibfnamefont{M.}~\bibnamefont{Golubitsky}} \bibnamefont{and}
  \bibinfo{author}{\bibfnamefont{D.~G.} \bibnamefont{Schaeffer}},
  \emph{\bibinfo{title}{Singularities and Groups in Bifurcation Theory}},
  vol.~\bibinfo{volume}{1} (\bibinfo{publisher}{Springer--Verlag},
  \bibinfo{address}{New York}, \bibinfo{year}{1985}).

\bibitem[{\citenamefont{Holmes et~al.}(1996)\citenamefont{Holmes, Lumley, and
  Berkooz}}]{Holmes:1996}
\bibinfo{author}{\bibfnamefont{P.}~\bibnamefont{Holmes}},
  \bibinfo{author}{\bibfnamefont{J.~L.} \bibnamefont{Lumley}},
  \bibnamefont{and} \bibinfo{author}{\bibfnamefont{G.}~\bibnamefont{Berkooz}},
  \emph{\bibinfo{title}{Turbulence, Coherent Structures, Dynamical Systems and
  Symmetry}} (\bibinfo{publisher}{Cambridge University Press},
  \bibinfo{address}{Cambridge}, \bibinfo{year}{1996}).

\bibitem[{\citenamefont{Ball and Dewar}(2000)}]{Ball:2000}
\bibinfo{author}{\bibfnamefont{R.}~\bibnamefont{Ball}} \bibnamefont{and}
  \bibinfo{author}{\bibfnamefont{R.~L.} \bibnamefont{Dewar}},
  \bibinfo{journal}{Phys. Rev. Lett.} \textbf{\bibinfo{volume}{84}},
  \bibinfo{pages}{3077} (\bibinfo{year}{2000}).

\bibitem[{\citenamefont{James}(1987)}]{James:1987}
\bibinfo{author}{\bibfnamefont{I.~N.} \bibnamefont{James}},
  \bibinfo{journal}{Journal of the Atmospheric Sciences}
  \textbf{\bibinfo{volume}{44}}, \bibinfo{pages}{3710} (\bibinfo{year}{1987}).

\bibitem[{\citenamefont{Strauss}(1977)}]{Strauss:1977}
\bibinfo{author}{\bibfnamefont{H.~R.} \bibnamefont{Strauss}},
  \bibinfo{journal}{ Phys. Fluids} \textbf{\bibinfo{volume}{20}},
  \bibinfo{pages}{1354} (\bibinfo{year}{1977}).

\bibitem[{\citenamefont{Strauss}(1980)}]{Strauss:1980}
\bibinfo{author}{\bibfnamefont{H.~R.} \bibnamefont{Strauss}},
  \bibinfo{journal}{Plasma Physics} \textbf{\bibinfo{volume}{22}},
  \bibinfo{pages}{733} (\bibinfo{year}{1980}).

\bibitem[{\citenamefont{Braginskii}(1965)}]{Leontovich:1965}
\bibinfo{author}{\bibfnamefont{S.~I.} \bibnamefont{Braginskii}}, in
  \emph{\bibinfo{booktitle}{Reviews of Plasma Physics}}, edited by
  \bibinfo{editor}{\bibfnamefont{M.~A.} \bibnamefont{Leontovich}}
  (\bibinfo{publisher}{Consultants Bureau}, \bibinfo{address}{New York},
  \bibinfo{year}{1965}), vol.~\bibinfo{volume}{1}.

\bibitem[{\citenamefont{Sugama and Horton}(1994{\natexlab{a}})}]{Sugama:1994a}
\bibinfo{author}{\bibfnamefont{H.}~\bibnamefont{Sugama}} \bibnamefont{and}
  \bibinfo{author}{\bibfnamefont{W.}~\bibnamefont{Horton}},
  \bibinfo{journal}{Phys. Plasmas} \textbf{\bibinfo{volume}{1}},
  \bibinfo{pages}{345} (\bibinfo{year}{1994}{\natexlab{a}}).

\bibitem[{\citenamefont{Ida et~al.}(1990)\citenamefont{Ida, Hidekuma, Miura,
  Fujita, Mori, Hoshino, Suzuki, Yamauchi, and {JFT-2M Group}}}]{Ida:1990}
\bibinfo{author}{\bibfnamefont{K.}~\bibnamefont{Ida}},
  \bibinfo{author}{\bibfnamefont{S.}~\bibnamefont{Hidekuma}},
  \bibinfo{author}{\bibfnamefont{Y.}~\bibnamefont{Miura}},
  \bibinfo{author}{\bibfnamefont{T.}~\bibnamefont{Fujita}},
  \bibinfo{author}{\bibfnamefont{M.}~\bibnamefont{Mori}},
  \bibinfo{author}{\bibfnamefont{K.}~\bibnamefont{Hoshino}},
  \bibinfo{author}{\bibfnamefont{N.}~\bibnamefont{Suzuki}},
  \bibinfo{author}{\bibfnamefont{T.}~\bibnamefont{Yamauchi}}, \bibnamefont{and}
  \bibinfo{author}{\bibnamefont{{JFT-2M Group}}}, \bibinfo{journal}{Phys. Rev.
  Lett.} \textbf{\bibinfo{volume}{65}}, \bibinfo{pages}{1364}
  (\bibinfo{year}{1990}).

\bibitem[{\citenamefont{Thomas et~al.}(1998)\citenamefont{Thomas, Groebner,
  Burrell, Osborne, and Carlstrom}}]{Thomas:1998a}
\bibinfo{author}{\bibfnamefont{D.}~\bibnamefont{Thomas}},
  \bibinfo{author}{\bibfnamefont{R.}~\bibnamefont{Groebner}},
  \bibinfo{author}{\bibfnamefont{K.}~\bibnamefont{Burrell}},
  \bibinfo{author}{\bibfnamefont{T.}~\bibnamefont{Osborne}}, \bibnamefont{and}
  \bibinfo{author}{\bibfnamefont{T.}~\bibnamefont{Carlstrom}},
  \bibinfo{journal}{Plasma Phys. Control. Fusion}
  \textbf{\bibinfo{volume}{40}}, \bibinfo{pages}{707} (\bibinfo{year}{1998}).

\bibitem[{\citenamefont{Igitkhanov et~al.}(1998)\citenamefont{Igitkhanov,
  Janeschitz, Pacher, Sugihara, Pacher, Post, Solano, Lingertat, Loarte,
  Osborne et~al.}}]{Igitkhanov:1998}
\bibinfo{author}{\bibfnamefont{Y.}~\bibnamefont{Igitkhanov}},
  \bibinfo{author}{\bibfnamefont{G.}~\bibnamefont{Janeschitz}},
  \bibinfo{author}{\bibfnamefont{G.~W.} \bibnamefont{Pacher}},
  \bibinfo{author}{\bibfnamefont{M.}~\bibnamefont{Sugihara}},
  \bibinfo{author}{\bibfnamefont{H.~D.} \bibnamefont{Pacher}},
  \bibinfo{author}{\bibfnamefont{D.~E.} \bibnamefont{Post}},
  \bibinfo{author}{\bibfnamefont{E.}~\bibnamefont{Solano}},
  \bibinfo{author}{\bibfnamefont{J.}~\bibnamefont{Lingertat}},
  \bibinfo{author}{\bibfnamefont{A.}~\bibnamefont{Loarte}},
  \bibinfo{author}{\bibfnamefont{T.}~\bibnamefont{Osborne}},
  \bibnamefont{et~al.}, \bibinfo{journal}{Plasma Phys. Control. Fusion}
  \textbf{\bibinfo{volume}{40}}, \bibinfo{pages}{837} (\bibinfo{year}{1998}).

\bibitem[{\citenamefont{Shats}(1999)}]{Shats:1999}
\bibinfo{author}{\bibfnamefont{M.~G.} \bibnamefont{Shats}},
  \bibinfo{journal}{Plasma Phys. Control. Fusion}
  \textbf{\bibinfo{volume}{41}}, \bibinfo{pages}{1357} (\bibinfo{year}{1999}).

\bibitem[{\citenamefont{Bak et~al.}(2001)\citenamefont{Bak, Asakura, Miura,
  Nakano, and Yoshino}}]{PEBak:2001}
\bibinfo{author}{\bibfnamefont{P.~E.} \bibnamefont{Bak}},
  \bibinfo{author}{\bibfnamefont{N.}~\bibnamefont{Asakura}},
  \bibinfo{author}{\bibfnamefont{Y.}~\bibnamefont{Miura}},
  \bibinfo{author}{\bibfnamefont{T.}~\bibnamefont{Nakano}}, \bibnamefont{and}
  \bibinfo{author}{\bibfnamefont{R.}~\bibnamefont{Yoshino}},
  \bibinfo{journal}{Phys. Plasmas} \textbf{\bibinfo{volume}{8}},
  \bibinfo{pages}{1248} (\bibinfo{year}{2001}).

\bibitem[{\citenamefont{Connor}(1998)}]{Connor:1998}
\bibinfo{author}{\bibfnamefont{J.~W.} \bibnamefont{Connor}},
  \bibinfo{journal}{Plasma Phys. Control. Fusion}
  \textbf{\bibinfo{volume}{40}}, \bibinfo{pages}{531} (\bibinfo{year}{1998}).

\bibitem[{\citenamefont{Suttrop}(2000)}]{Suttrop:2000}
\bibinfo{author}{\bibfnamefont{W.}~\bibnamefont{Suttrop}},
  \bibinfo{journal}{Plasma Phys. Control. Fusion}
  \textbf{\bibinfo{volume}{42}}, \bibinfo{pages}{A1} (\bibinfo{year}{2000}).

\bibitem[{\citenamefont{Burrell et~al.}(2001)\citenamefont{Burrell, Austin,
  Brennan, DeBoo, Doyle, Fenzi, Fuchs, Gohil, Greenfield, Groebner
  et~al.}}]{Burrell:2001}
\bibinfo{author}{\bibfnamefont{K.~H.} \bibnamefont{Burrell}},
  \bibinfo{author}{\bibfnamefont{M.~E.} \bibnamefont{Austin}},
  \bibinfo{author}{\bibfnamefont{D.~P.} \bibnamefont{Brennan}},
  \bibinfo{author}{\bibfnamefont{J.~C.} \bibnamefont{DeBoo}},
  \bibinfo{author}{\bibfnamefont{E.~J.} \bibnamefont{Doyle}},
  \bibinfo{author}{\bibfnamefont{C.}~\bibnamefont{Fenzi}},
  \bibinfo{author}{\bibfnamefont{C.}~\bibnamefont{Fuchs}},
  \bibinfo{author}{\bibfnamefont{P.}~\bibnamefont{Gohil}},
  \bibinfo{author}{\bibfnamefont{C.~M.} \bibnamefont{Greenfield}},
  \bibinfo{author}{\bibfnamefont{R.~J.} \bibnamefont{Groebner}},
  \bibnamefont{et~al.}, \bibinfo{journal}{Phys. Plasmas}
  \textbf{\bibinfo{volume}{8}}, \bibinfo{pages}{2153} (\bibinfo{year}{2001}).

\bibitem[{\citenamefont{Bell et~al.}(1998)\citenamefont{Bell, Levinton, Batha,
  Synakowski, and Zarnstorff}}]{Bell:1998}
\bibinfo{author}{\bibfnamefont{R.~E.} \bibnamefont{Bell}},
  \bibinfo{author}{\bibfnamefont{F.~M.} \bibnamefont{Levinton}},
  \bibinfo{author}{\bibfnamefont{S.~H.} \bibnamefont{Batha}},
  \bibinfo{author}{\bibfnamefont{E.~J.} \bibnamefont{Synakowski}},
  \bibnamefont{and} \bibinfo{author}{\bibfnamefont{M.~C.}
  \bibnamefont{Zarnstorff}}, \bibinfo{journal}{Phys. Rev. Lett.}
  \textbf{\bibinfo{volume}{81}}, \bibinfo{pages}{1429} (\bibinfo{year}{1998}).

\bibitem[{\citenamefont{Solomon}(2001)}]{Solomon:2001}
\bibinfo{author}{\bibfnamefont{W.}~\bibnamefont{Solomon}}
  \bibinfo{note}{private communication}.

\bibitem[{\citenamefont{Zohm et~al.}(1995)\citenamefont{Zohm, Suttrop, Buchl,
  deBlank, Gruber, Kallenbach, Mertens, Ryter, and Schittenhelm}}]{Zohm:1995}
\bibinfo{author}{\bibfnamefont{H.}~\bibnamefont{Zohm}},
  \bibinfo{author}{\bibfnamefont{W.}~\bibnamefont{Suttrop}},
  \bibinfo{author}{\bibfnamefont{K.}~\bibnamefont{Buchl}},
  \bibinfo{author}{\bibfnamefont{H.~J.} \bibnamefont{deBlank}},
  \bibinfo{author}{\bibfnamefont{O.}~\bibnamefont{Gruber}},
  \bibinfo{author}{\bibfnamefont{A.}~\bibnamefont{Kallenbach}},
  \bibinfo{author}{\bibfnamefont{V.}~\bibnamefont{Mertens}},
  \bibinfo{author}{\bibfnamefont{F.}~\bibnamefont{Ryter}}, \bibnamefont{and}
  \bibinfo{author}{\bibfnamefont{M.}~\bibnamefont{Schittenhelm}},
  \bibinfo{journal}{Plasma Phys. Control. Fusion}
  \textbf{\bibinfo{volume}{37}}, \bibinfo{pages}{437} (\bibinfo{year}{1995}).

\bibitem[{\citenamefont{Ryter et~al.}(1998)\citenamefont{Ryter, Suttrop,
  Br{\"u}sehaber, Kaufmann, Mertens, Murmann, Peeters, Stober, Schweinzer, Zohm
  et~al.}}]{Ryter:1998}
\bibinfo{author}{\bibfnamefont{F.}~\bibnamefont{Ryter}},
  \bibinfo{author}{\bibfnamefont{W.}~\bibnamefont{Suttrop}},
  \bibinfo{author}{\bibfnamefont{B.}~\bibnamefont{Br{\"u}sehaber}},
  \bibinfo{author}{\bibfnamefont{M.}~\bibnamefont{Kaufmann}},
  \bibinfo{author}{\bibfnamefont{V.}~\bibnamefont{Mertens}},
  \bibinfo{author}{\bibfnamefont{H.}~\bibnamefont{Murmann}},
  \bibinfo{author}{\bibfnamefont{A.~G.} \bibnamefont{Peeters}},
  \bibinfo{author}{\bibfnamefont{J.}~\bibnamefont{Stober}},
  \bibinfo{author}{\bibfnamefont{J.}~\bibnamefont{Schweinzer}},
  \bibinfo{author}{\bibfnamefont{H.}~\bibnamefont{Zohm}}, \bibnamefont{et~al.},
  \bibinfo{journal}{Plasma Phys. Control. Fusion}
  \textbf{\bibinfo{volume}{40}}, \bibinfo{pages}{725} (\bibinfo{year}{1998}).

\bibitem[{\citenamefont{Hubbard et~al.}(1998)\citenamefont{Hubbard, Boivin,
  Drake, Greenwald, In, Irby, Rogers, and Snipes}}]{Hubbard:1998}
\bibinfo{author}{\bibfnamefont{A.~E.} \bibnamefont{Hubbard}},
  \bibinfo{author}{\bibfnamefont{R.~L.} \bibnamefont{Boivin}},
  \bibinfo{author}{\bibfnamefont{J.~F.} \bibnamefont{Drake}},
  \bibinfo{author}{\bibfnamefont{M.}~\bibnamefont{Greenwald}},
  \bibinfo{author}{\bibfnamefont{Y.}~\bibnamefont{In}},
  \bibinfo{author}{\bibfnamefont{J.~H.} \bibnamefont{Irby}},
  \bibinfo{author}{\bibfnamefont{B.~N.} \bibnamefont{Rogers}},
  \bibnamefont{and} \bibinfo{author}{\bibfnamefont{J.~A.}
  \bibnamefont{Snipes}}, \bibinfo{journal}{Plasma Phys. Control. Fusion}
  \textbf{\bibinfo{volume}{40}}, \bibinfo{pages}{689} (\bibinfo{year}{1998}).

\bibitem[{\citenamefont{Moyer et~al.}(1999)\citenamefont{Moyer, Rhodes, Rettig,
  Doyle, Burrell, Cuthbertson, Groebner, Kim, Leonard, Maingi
  et~al.}}]{Moyer:1999}
\bibinfo{author}{\bibfnamefont{R.~A.} \bibnamefont{Moyer}},
  \bibinfo{author}{\bibfnamefont{T.~L.} \bibnamefont{Rhodes}},
  \bibinfo{author}{\bibfnamefont{C.~L.} \bibnamefont{Rettig}},
  \bibinfo{author}{\bibfnamefont{E.~J.} \bibnamefont{Doyle}},
  \bibinfo{author}{\bibfnamefont{K.~H.} \bibnamefont{Burrell}},
  \bibinfo{author}{\bibfnamefont{J.}~\bibnamefont{Cuthbertson}},
  \bibinfo{author}{\bibfnamefont{R.~J.} \bibnamefont{Groebner}},
  \bibinfo{author}{\bibfnamefont{K.~W.} \bibnamefont{Kim}},
  \bibinfo{author}{\bibfnamefont{A.~W.} \bibnamefont{Leonard}},
  \bibinfo{author}{\bibfnamefont{R.}~\bibnamefont{Maingi}},
  \bibnamefont{et~al.}, \bibinfo{journal}{Plasma Phys. Control. Fusion}
  \textbf{\bibinfo{volume}{41}}, \bibinfo{pages}{243} (\bibinfo{year}{1999}).

\bibitem[{\citenamefont{Dahi et~al.}(1998)\citenamefont{Dahi, Talmadge, and
  Shohet}}]{Dahi:1998}
\bibinfo{author}{\bibfnamefont{H.}~\bibnamefont{Dahi}},
  \bibinfo{author}{\bibfnamefont{J.~N.} \bibnamefont{Talmadge}},
  \bibnamefont{and} \bibinfo{author}{\bibfnamefont{J.~L.}
  \bibnamefont{Shohet}}, \bibinfo{journal}{Phys. Rev. Lett.}
  \textbf{\bibinfo{volume}{80}}, \bibinfo{pages}{3976} (\bibinfo{year}{1998}).

\bibitem[{\citenamefont{Shats et~al.}(1997)\citenamefont{Shats, Rudakov,
  Boswell, and Borg}}]{Shats:1997a}
\bibinfo{author}{\bibfnamefont{M.~G.} \bibnamefont{Shats}},
  \bibinfo{author}{\bibfnamefont{D.~L.} \bibnamefont{Rudakov}},
  \bibinfo{author}{\bibfnamefont{R.~W.} \bibnamefont{Boswell}},
  \bibnamefont{and} \bibinfo{author}{\bibfnamefont{G.~G.} \bibnamefont{Borg}},
  \bibinfo{journal}{Phys. Plasmas} \textbf{\bibinfo{volume}{4}},
  \bibinfo{pages}{3629} (\bibinfo{year}{1997}).

\bibitem[{\citenamefont{Hugill et~al.}(1998)\citenamefont{Hugill, Broomhead,
  and Barratt}}]{Hugill:1998}
\bibinfo{author}{\bibfnamefont{J.}~\bibnamefont{Hugill}},
  \bibinfo{author}{\bibfnamefont{D.~S.} \bibnamefont{Broomhead}},
  \bibnamefont{and} \bibinfo{author}{\bibfnamefont{M.}~\bibnamefont{Barratt}},
  \bibinfo{journal}{ICPP\&25th EPS Conf. on Contr. Fusion and Plasma Phys. 
  ECA} \textbf{\bibinfo{volume}{22C}},
  \bibinfo{pages}{2318} (\bibinfo{year}{1998}).

\bibitem[{\citenamefont{Hirsch et~al.}(1998)\citenamefont{Hirsch, Amadeo,
  Anton, Baldzuhn, Brakel, Bleuel, Fiedler, Geist, Grigull, Hartfuss
  et~al.}}]{Hirsch:1998}
\bibinfo{author}{\bibfnamefont{M.}~\bibnamefont{Hirsch}},
  \bibinfo{author}{\bibfnamefont{P.}~\bibnamefont{Amadeo}},
  \bibinfo{author}{\bibfnamefont{M.}~\bibnamefont{Anton}},
  \bibinfo{author}{\bibfnamefont{J.}~\bibnamefont{Baldzuhn}},
  \bibinfo{author}{\bibfnamefont{R.}~\bibnamefont{Brakel}},
  \bibinfo{author}{\bibfnamefont{J.}~\bibnamefont{Bleuel}},
  \bibinfo{author}{\bibfnamefont{S.}~\bibnamefont{Fiedler}},
  \bibinfo{author}{\bibfnamefont{T.}~\bibnamefont{Geist}},
  \bibinfo{author}{\bibfnamefont{P.}~\bibnamefont{Grigull}},
  \bibinfo{author}{\bibfnamefont{H.~J.} \bibnamefont{Hartfuss}},
  \bibnamefont{et~al.}, \bibinfo{journal}{Plasma Phys. Control. Fusion}
  \textbf{\bibinfo{volume}{40}}, \bibinfo{pages}{631} (\bibinfo{year}{1998}).

\bibitem[{\citenamefont{Hirsch et~al.}(2000)\citenamefont{Hirsch, Grigull,
  Kisslinger, McCormick, Anton, Baldzuhn, Fiedler, Fuchs, Geiger, Giannone
  et~al.}}]{Hirsch:2000}
\bibinfo{author}{\bibfnamefont{M.}~\bibnamefont{Hirsch}},
  \bibinfo{author}{\bibfnamefont{H.}~\bibnamefont{Grigull},
  \bibfnamefont{P.~Wobig}},
  \bibinfo{author}{\bibfnamefont{J.}~\bibnamefont{Kisslinger}},
  \bibinfo{author}{\bibfnamefont{K.}~\bibnamefont{McCormick}},
  \bibinfo{author}{\bibfnamefont{M.}~\bibnamefont{Anton}},
  \bibinfo{author}{\bibfnamefont{J.}~\bibnamefont{Baldzuhn}},
  \bibinfo{author}{\bibfnamefont{S.}~\bibnamefont{Fiedler}},
  \bibinfo{author}{\bibfnamefont{C.}~\bibnamefont{Fuchs}},
  \bibinfo{author}{\bibfnamefont{J.}~\bibnamefont{Geiger}},
  \bibinfo{author}{\bibfnamefont{H.-J.} \bibnamefont{Giannone},
  \bibfnamefont{L.~H-J~Hartfuss}}, \bibnamefont{et~al.},
  \bibinfo{journal}{Plasma Phys. Control. Fusion}
  \textbf{\bibinfo{volume}{42}}, \bibinfo{pages}{A231} (\bibinfo{year}{2000}).

\bibitem[{\citenamefont{Shats and Rudakov}(1997)}]{Shats:1997b}
\bibinfo{author}{\bibfnamefont{M.~G.} \bibnamefont{Shats}} \bibnamefont{and}
  \bibinfo{author}{\bibfnamefont{D.~L.} \bibnamefont{Rudakov}},
  \bibinfo{journal}{Phys. Rev. Lett.} \textbf{\bibinfo{volume}{79}},
  \bibinfo{pages}{2690} (\bibinfo{year}{1997}).

\bibitem[{\citenamefont{Stix}(1973)}]{Stix:1973}
\bibinfo{author}{\bibfnamefont{T.~H.} \bibnamefont{Stix}},
  \bibinfo{journal}{Phys. Fluids} \textbf{\bibinfo{volume}{16}},
  \bibinfo{pages}{1260} (\bibinfo{year}{1973}).

\bibitem[{\citenamefont{Carreras et~al.}(1987)\citenamefont{Carreras, Garcia,
  and Diamond}}]{Carreras:1987}
\bibinfo{author}{\bibfnamefont{B.~A.} \bibnamefont{Carreras}},
  \bibinfo{author}{\bibfnamefont{L.}~\bibnamefont{Garcia}}, \bibnamefont{and}
  \bibinfo{author}{\bibfnamefont{P.~H.} \bibnamefont{Diamond}},
  \bibinfo{journal}{Phys. Fluids} \textbf{\bibinfo{volume}{30}},
  \bibinfo{pages}{1388} (\bibinfo{year}{1987}).

\bibitem[{\citenamefont{Sugama and Wakatani}(1988)}]{Sugama:1988}
\bibinfo{author}{\bibfnamefont{H.}~\bibnamefont{Sugama}} \bibnamefont{and}
  \bibinfo{author}{\bibfnamefont{M.}~\bibnamefont{Wakatani}},
  \bibinfo{journal}{J. Phys. Soc. Japan} \textbf{\bibinfo{volume}{57}},
  \bibinfo{pages}{2010} (\bibinfo{year}{1988}).

\bibitem[{\citenamefont{Sugama and Horton}(1994{\natexlab{b}})}]{Sugama:1994b}
\bibinfo{author}{\bibfnamefont{H.}~\bibnamefont{Sugama}} \bibnamefont{and}
  \bibinfo{author}{\bibfnamefont{W.}~\bibnamefont{Horton}},
  \bibinfo{journal}{Phys. Plasmas} \textbf{\bibinfo{volume}{1}},
  \bibinfo{pages}{2220} (\bibinfo{year}{1994}{\natexlab{b}}).

\end{thebibliography}

\clearpage
\appendix{\bf Appendix\label{appendix}}

\begin{figure} 
\hspace*{0cm}\includegraphics[scale=0.8]{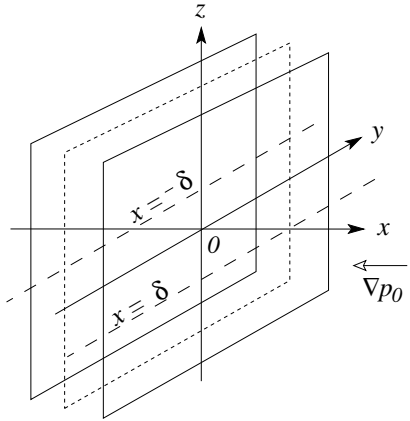}	
\caption{Simple slab geometry is assumed. The plasma edge region is $-\delta<x<\delta$,
with $x=\delta$ at the plasma surface.
$\nabla p_0 <0$ is the $y,z$-averaged pressure gradient. }\label{figure8}
\end{figure}

Reduced MHD fluid equations in tokamak and stellarator geometries
were originally derived by Strauss \cite{Strauss:1977,Strauss:1980}. 
In the electrostatic approximation, a damped MHD fluid may be described by 
the following momentum and pressure convection equations:

\begin{align}
\rho\frac{d{\rm\mathbf v}}{dt}  & = -\nabla p + 
              {\rm\mathbf J} \times {\rm\mathbf B}
              + \mu\nabla_\perp^2 {\rm\mathbf v}
              + \Omega^\prime \tilde{p} \hat{x}
              - \rho\nu\left( {\rm\mathbf v} - V\left(x\right)\hat{y} \right)
\label{momentum}\\
\frac{dp}{dt} & = \chi\nabla_\perp^2 p ,\label{pressure}
\end{align}
where $ d/dt = \partial/\partial t + {\rm\mathbf v} \cdot \nabla $,  
together with the incompressibility condition $ \nabla \cdot {\rm\mathbf v} = 0 $
and the resistive Ohm's law 
$ {\rm\mathbf E} + {\rm\mathbf v}\times{\rm\mathbf B} = \eta{\rm\mathbf J} $. 
The symbols and notation are explained in table~\ref{table1}. 
The curl of Eq. \ref{momentum} yields a vorticity equation, which is sometimes
preferred in two-dimensional fluid dynamics, but we have used the momentum 
form because it is more transparent physically and has a simpler correspondence
to the kinetic energy. 
An infinite slab configuration is used for simplicity and generality, as was
also assumed in \cite{Stix:1973} for a drift-kinetic treatment of plasma relaxation.
It is sketched in Fig.~\ref{figure8}, 
where the region $ -\delta < x < \delta $ can be taken to represent a region 
at the edge or within a confined plasma where a transport barrier evolves.

The last term on the right hand side of Eq. \ref{momentum} 
removes the nonlinear shear-flow reversal 
symmetry of the system under $ v_x(x,y,t) \rightarrow v_x(x,-y,t)  $,
$ v_y(x,y,t) \rightarrow -v_y(x,-y,t)  $,
$ \tilde{p}(x,y,t) \rightarrow \tilde{p}(x,-y,t)  $,
$ V(x) \rightarrow V(x)  $.
Similar equations, without the symmetry-breaking term in the momentum balance,
 have been used by several authors  
\cite{Carreras:1987,Sugama:1988,Sugama:1994a,Sugama:1994b,Sugama:1995} 
as a basis for studying
resistive turbulence--flow interactions. 
The symmetry-breaking term was introduced in \cite{Sugama:1994a}, but only 
\textit{a posteriori} as an adjunct in an equation for the background poloidal flow. 
Here we introduce it at the outset. It models the friction force acting 
between the single-fluid plasma 
velocity $\mathbf{v} $ and an assumed external poloidal flow $V\hat{y}$. 
Although $V\hat{y}$ may be  described for convenience as an external velocity, 
the term 
represents any asymmetric shear-inducing mechanism, such as friction with 
neutrals, non-ambipolar ion orbit losses,
 or neoclassical effects not included in the slab model. 

\begin{figure} 
\hspace*{0cm}\includegraphics[scale=0.8]{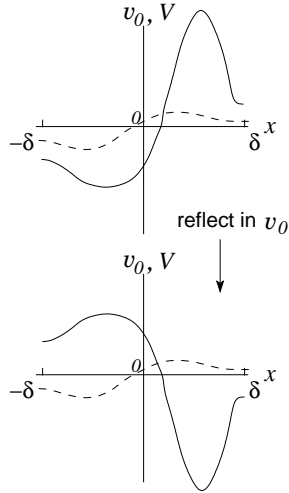}
\caption{\label{figure9}Without the friction force the system is invariant under the
transformation $v_0(x,t) \rightarrow - v_0(x, t)$ (solid line), 
$V (x) \rightarrow  V (x)$ (dashed line). When the friction coefficient
$\nu \neq 0$ the symmetry is broken.}
\end{figure}

The symmetry operation on $v_0(x)\equiv \langle v_y\rangle(x)$ 
and $V(x)$ is sketched in Fig. \ref{figure9}.
We are working in the frame in which there is no electrostatic potential
difference across the slab. That is, it is assumed that we have made a
Galilean transformation to the frame in which the spatial average of
$v_0$ across the slab is zero. For simplicity we also assume that the
spatial average of $V$ is zero.

Equations \ref{momentum} and \ref{pressure} are not intended to express a 
detailed fluid description of a plasma, but are intended instead to represent 
a qualitatively authentic, semi-empirical model for the essential generation and 
loss processes 
that give rise to the turbulence--shear flow interactions that we have schematized 
in Fig. \ref{fig1} as the plasma turbulence governor. 
The dynamical system Eqs \ref{dp}--\ref{dv} can be derived from Eqs \ref{momentum}
and \ref{pressure}, following the spatial averaging 
procedure implicit in \cite{Sugama:1995}. 

First of all, the dynamics of the mean flow $v_0=\langle v_y\rangle$ are
 extracted from the first moment ($v_y\hat{y}$) of 
$\langle$ Eq. \ref{momentum} $\rangle$ as 
\begin{equation}\label{dv0dt}
\partial_tv_0-\mu\partial^2_xv_0
+\partial_x\langle\tilde{v}_x\tilde{v}_y\rangle = -\nu(v_0 - V),
\end{equation}
the energy moment of which gives the spatially averaged evolution of shear
 flow kinetic energy $F$:
\begin{equation}
\begin{split}
\frac{d}{dt}\left[\frac{1}{\delta}\int_{-\delta}^\delta\frac{dx}{2}\:v_0^2\right]
 &= 
- \frac{1}{\delta}\int_{-\delta}^\delta dx\:\left[\mu\left( \frac{dv_0}{dx}\right)^2
+ \nu\:v_0^2\right] \\
&\quad\quad + \frac{1}{\delta}
\int_{-\delta}^\delta dx\:\langle\tilde{v}_x\tilde{v}_y\rangle\frac{dv_0}{dx}\\
&\quad\quad\quad\quad+\frac{1}{\delta}
\int_{-\delta}^\delta dx\:\nu V v_0.
\end{split}
\label{dF0}
\end{equation}
This may be written as
\begin{equation}\label{shortF} 
\frac{dF}{dt} = -\epsilon_F + E_F + E_\varphi,
\end{equation}
where the definitions of $\epsilon_F$, $E_F$, 
and $E_\varphi$ correspond respectively to each term on the right hand side of  
Eq. \ref{dF0} and $F\equiv \frac{1}{\delta}\int_{-\delta}^\delta\frac{dx}{2}\:v_0^2$. 

Next, the second moment of Eq. \ref{momentum} gives the total rate of evolution of
$F$ and turbulent kinetic energy $N$: 

\begin{equation}\label{FandN}
\begin{split}
\frac{d}{dt}\left[\frac{1}{\delta}\int_{-\delta}^\delta\frac{dx}{2}
\left(v_0^2+\tilde{v}^2\right)\right]
 &=
\frac{1}{\delta}\int_{-\delta}^\delta dx\:
\Omega^\prime\frac{\left<\tilde{p}\,\tilde{v}_x\right>}{\rho}\\
 &\quad -
\frac{1}{\delta}\int_{-\delta}^\delta dx\,\left[
\frac{\eta}{\rho_m}\left<\tilde{J}_\parallel^2\right>
+ \mu\left<\left(\frac{\partial\tilde{v}_i}{\partial x_j}\right)^2\right>\right]\\
 &\quad\quad -
\frac{1}{\delta}
\int_{-\delta}^\delta dx\:\left[\mu\left(\frac{dv_0}{dx}\right)^2 + \nu\:v_0^2\right] \\
&\quad\quad\quad + 
\frac{1}{\delta}
\int_{-\delta}^\delta dx\:\nu V v_0,
\end{split}
\end{equation}
which may be expressed succinctly as
\begin{equation}\label{shortFN}
\frac{d}{dt}\left[F+N\right] = E_N-\epsilon_N-\epsilon_F + E_\varphi
\end{equation}
where $E_N$ and $\epsilon_N$ are defined by the first two terms on the 
right hand side of Eq. \ref{FandN} and 
$N\equiv \frac{1}{\delta}\int_{-\delta}^\delta\frac{dx}{2}\:\tilde{v}^2$ . 

Finally, the evolution of potential energy in the pressure gradient is defined from 
the $x$-moment of Eq. \ref{pressure}, assuming the cross-field thermal
transport $\chi\nabla_\perp^2$ can be neglected:
\begin{equation}\label{xmoment}
\frac{d}{dt}\left[\frac{1}{\delta}\int_{-\delta}^\delta dx\:
\left(-x\right)\Omega^\prime\frac{p_0}{\rho}\right]
 =
\frac{\langle\tilde{p}\tilde{v}_x\rangle|_{-\delta}^\delta}{\rho}\Omega^\prime
- \frac{1}{\delta}\int_{-\delta}^\delta dx\:
\frac{\langle\tilde{p}\tilde{v}_x\rangle}{\rho} \Omega^\prime,
\end{equation}
or
\begin{equation}
\frac{dP}{dt} =  E_P - E_N \label{shortP},
\end{equation}
with the input rate $E_P$ defined as the first term on the right hand side 
and $P\equiv \frac{1}{\delta}\int_{-\delta}^\delta dx\:
\left(-x\right)\Omega^\prime p_0/\rho$. 

The spatially averaged dynamical system thus consists of Eqs \ref{shortFN},
\ref{shortF}, and \ref{shortP}. 
For closure we follow \cite{Sugama:1995}, using
the approximations $p_0 (x) \simeq p_0(x=\delta) + xdp_0/dx$ and 
$v_0 (x) \simeq v_0(x=\delta) + xdv_0/dx$ 
for the background pressure and flow profiles and re-defining $P$ and $F$ as
the gradient terms alone. 
Approximations or expressions based on empirical arguments 
were given in \cite{Sugama:1995} for the rates in
 Eqs \ref{shortF}, \ref{shortFN}, and \ref{shortP}.
 The rates given in Eqs \ref{dp}--\ref{dv} are economized versions of 
those expressions, in the sense that simpler power laws were chosen if
this did not result in any qualitative changes to the singularity 
and stability structure of the system.  
The rationale is that for most of the rates we shall only learn from experiments
whether different powers apply, meanwhile  simple power laws 
give more transparent algebra. 
We approximate 
the energy transfer rate from the pressure gradient
simply as $E_N \simeq (\gamma/\varepsilon) PN$, and the energy transfer rate 
between the
turbulence and the shear flow, due to the Reynolds stress, as $E_F \simeq \alpha FN$. 
The power input through the boundary is defined as $E_p \equiv q/\varepsilon$. 
The two-timing coefficient $\varepsilon$ is related to the thermal
capacitance, and regulates
the contribution of the pressure gradient to the dynamics.
For the dissipative terms we take the turbulent energy dissipation rate
as $\epsilon_N \simeq \beta N^2$  and the shear flow energy damping rate 
as $\epsilon_F \simeq \mu (P, N)F$, assuming the viscous damping to be
dominant in $\epsilon_F $. The external shear flow driving rate is then
$E_\varphi \simeq \varphi F^{1/2}$, with $\varphi \simeq \delta \nu V$.
To obtain the evolution of the shear flow in terms of a velocity
gradient variable we re-define $v \equiv \pm F^{1/2}$. 
Eqs \ref{dp}--\ref{dv} ensue. 

\begin{table}[ht]
\begin{tabular}{ll}
$\mathbf{v} = \frac{1}{B_0}\mathbf{\hat{z}}\times\nabla\phi
=\mathbf{v}_0 + \tilde{\mathbf{v}}$\quad\quad&$\mathbf{E}\times\mathbf{B}$ flow velocity\\
$\mathbf{v}_0 = \langle\mathbf{v}\rangle$ & average background component\\
$\tilde{\rm\mathbf{v}}$ & fluctuating or turbulent component\\
$p = p_0 + \tilde{p}$ & plasma pressure \\
$p_0=\langle p\rangle$ & average background component\\
$\tilde{p}$ & fluctuating or turbulent component\\
$\rho$ & average mass density of ions, assumed constant\\
$\mu$ & ion viscosity coefficient\\
$B_0$ & magnetic field along the $z$ axis\\
$\eta$ & resistivity\\
$\nu$ & frictional damping coefficient\\
$\Omega^\prime\equiv d\Omega/dx>0$ &average field line curvature, assumed constant\\
$ \nabla_\perp^2$& $\partial^2_x+\partial^2_y$\\
$\nabla_\parallel $ & $ \partial_z+ \frac{x}{L_s}\partial_y$\\
$\chi$ &  cross-field thermal transport coefficient\\
$V$ & external flow\\
$\langle\cdots\rangle$ & average on $(y,z)$ plane 
\end{tabular}
\caption{\label{table1}Glossary of symbols, terms, and notation.}
\end{table}

\end{document}